\documentclass[letter,bibyear]{aa} 

\usepackage{natbib}
\usepackage{graphicx}
\usepackage{txfonts}

\newcommand{\jms}{J.~Mol.~Spectrosc.}   
\newcommand{\jmst}{J.~Mol.~Struct.}   

\newcommand{\kms}{km s$^{-1}$}

\newcommand{\once}{10$^{11}$\,cm$^{-2}$}

\begin{document}

\title{Discovery of benzyne, $o$-C$_6$H$_4$, in TMC-1 with the QUIJOTE$^1$ line survey\thanks{Based on observations carried out with the Yebes 40m telescope (projects 19A003, 20A014, 20D023, and 21A011). The 40m radio telescope at Yebes Observatory is operated by the Spanish Geographic Institute (IGN; Ministerio de Transportes, Movilidad y Agenda Urbana).}}

\author{
J.~Cernicharo\inst{1},
M.~Ag\'undez\inst{1},
R.~I.~Kaiser\inst{2},
C.~Cabezas\inst{1},
B.~Tercero\inst{3,4},
N.~Marcelino\inst{1},
J.~R.~Pardo\inst{1},
P.~de~Vicente\inst{3}
}

\institute{Grupo de Astrof\'isica Molecular, Instituto de F\'isica Fundamental (IFF-CSIC),
C/ Serrano 121, 28006 Madrid, Spain\\ \email jose.cernicharo@csic.es
\and Department of Chemistry, University of Hawaii at Manoa, Honolulu, HI 96822, USA
\and Centro de Desarrollos Tecnol\'ogicos, Observatorio de Yebes (IGN), 19141 Yebes, Guadalajara, Spain
\and Observatorio Astron\'omico Nacional (OAN, IGN), Madrid, Spain
}

\date{Received; accepted}
\abstract{
We report the detection, for the first time in space, of a new 
non-functionalised hydrocarbon cycle in the direction of TMC-1:
$o$-C$_6$H$_4$ (ortho-benzyne). We derive a column density for this hydrocarbon cycle of 
(5.0$\pm$1.0)$\times$10$^{11}$ cm$^{-2}$. The abundance of this species is around 30 times lower than that of
cyclopentadiene and indene. 
We compare the abundance of benzyne with that of other pure hydrocarbons, 
cycles or chains, and
find that it could be formed from neutral-radical reactions such as C$_2$H + 
CH$_2$CHCCH and C + C$_5$H$_5$, and possibly through C$_4$H + C$_2$H$_4$, 
C$_3$H + CH$_2$CCH$_2$, and C$_3$H$_2$+C$_3$H$_3$. Hence, the rich content of 
hydrocarbon cycles observed in TMC-1 could arise through a bottom-up scenario
involving reactions of a few 
radicals with the abundant hydrocarbons recently revealed by the 
QUIJOTE line survey.
}

\titlerunning{Benzyne in TMC-1}
\authorrunning{Cernicharo et al.}

\maketitle

\section{Introduction}
The recent discovery with the QUIJOTE\footnote{\textbf{Q}-band \textbf{U}ltrasensitive \textbf{I}nspection \textbf{J}ourney 
to the \textbf{O}bscure \textbf{T}MC-1 \textbf{E}nvironment} 
line survey of abundant hydrocarbons such as the propargyl radical (CH$_2$CCH), vinylacetylene 
(CH$_2$CHCCH), and ethynylallene (H$_2$CCCHCCH) and cyclic 
hydrocarbons such as cyclopentadiene ($c$-C$_5$H$_6$) and indene (c-C$_9$H$_8$) in the cold dark 
cloud TMC-1 \citep{Agundez2021a,Cernicharo2021a,Cernicharo2021b,Cernicharo2021c,Burkhardt2021},
as well as the detection of their cyano derivatives 
\citep{McGuire2018,McGuire2021} in the same source, points to a new and 
interesting chemistry that had been
completely unforeseen by the most sophisticated state-of-the-art chemical models. 
The detection
of indene, the first non-functionalised polycyclic aromatic hydrocarbons (PAHs) detected in space, 
in the unexpected environment
of a cold starless core is surprising and introduces new challenges to our
understanding of the place were these PAHs are formed in space. 

Many efforts have been devoted over the last 40 years to understanding the 
chemical processes that lead to the formation of these molecular species (see e.g. 
\citealt{Jones2011,Parker2012,Joblin2018}). 
Circumstellar envelopes around carbon-rich asymptotic giant branch (AGB) stars have been suggested 
as the factories of PAHs \citep{Cherchneff1992}. However, PAHs are only detected in 
well-UV-illuminated regions of the interstellar medium. 
The detection of benzene in the carbon-rich protoplanetary nebula CRL\,618 \citep{Cernicharo2001}
suggests a bottom-up approach in which the small hydrocarbons that formed during the AGB phase, such 
as C$_2$H$_2$ and C$_2$H$_4$, interact with the UV radiation produced by the star 
in its evolution to the white
dwarf phase \citep{Woods2002,Cernicharo2004}. Other hypotheses involve the processing of dust 
grains around evolved stars, either through UV photons \citep{Pilleri2015} or by chemical 
processes \citep{Martinez2020}.
Hence, it has been surprising to see that pure PAHs and their cyanide derivatives have been found in TMC-1, which is well protected against UV radiation 
\citep{Cernicharo2021c,Burkhardt2021,McGuire2018,McGuire2021}. It is unlikely that these molecules 
arise from a reservoir existing since the early stages of the cloud because these relatively 
small PAHs would not have survived the diffuse cloud stage. 
An in situ formation mechanism for benzene, cyclopentadiene, indene, and naphthalene must involve abundant 
hydrocarbons containing from two to five carbon atoms \citep{Kaiser2021}. 
Moreover, some of these
species should allow an easy cyclisation in bimolecular reactions to efficiently form the first aromatic 
ring -- benzene ($c$-C$_6$H$_6$), 
benzyne ($c$-C$_6$H$_4$), the phenyl radical ($c$-C$_6$H$_5$), or any other species -- from 
which larger PAHs can grow 
\citep{Jones2011,Parker2012,McCabe2020,Doddipatla2021}.
The progargyl radical and the closed-shell hydrocarbons methyl-,
vinyl-, and ethynylallene have been detected in TMC-1 with high abundances 
\citep{Agundez2021a,Cernicharo2021b,
Cernicharo2021c}. Detecting all possible products of the reactions between these hydrocarbons
and radicals is a mandatory step towards understand the chemistry of PAHs in cold dark clouds.

\begin{figure}[]
\centering
\includegraphics[scale=0.17,angle=0]{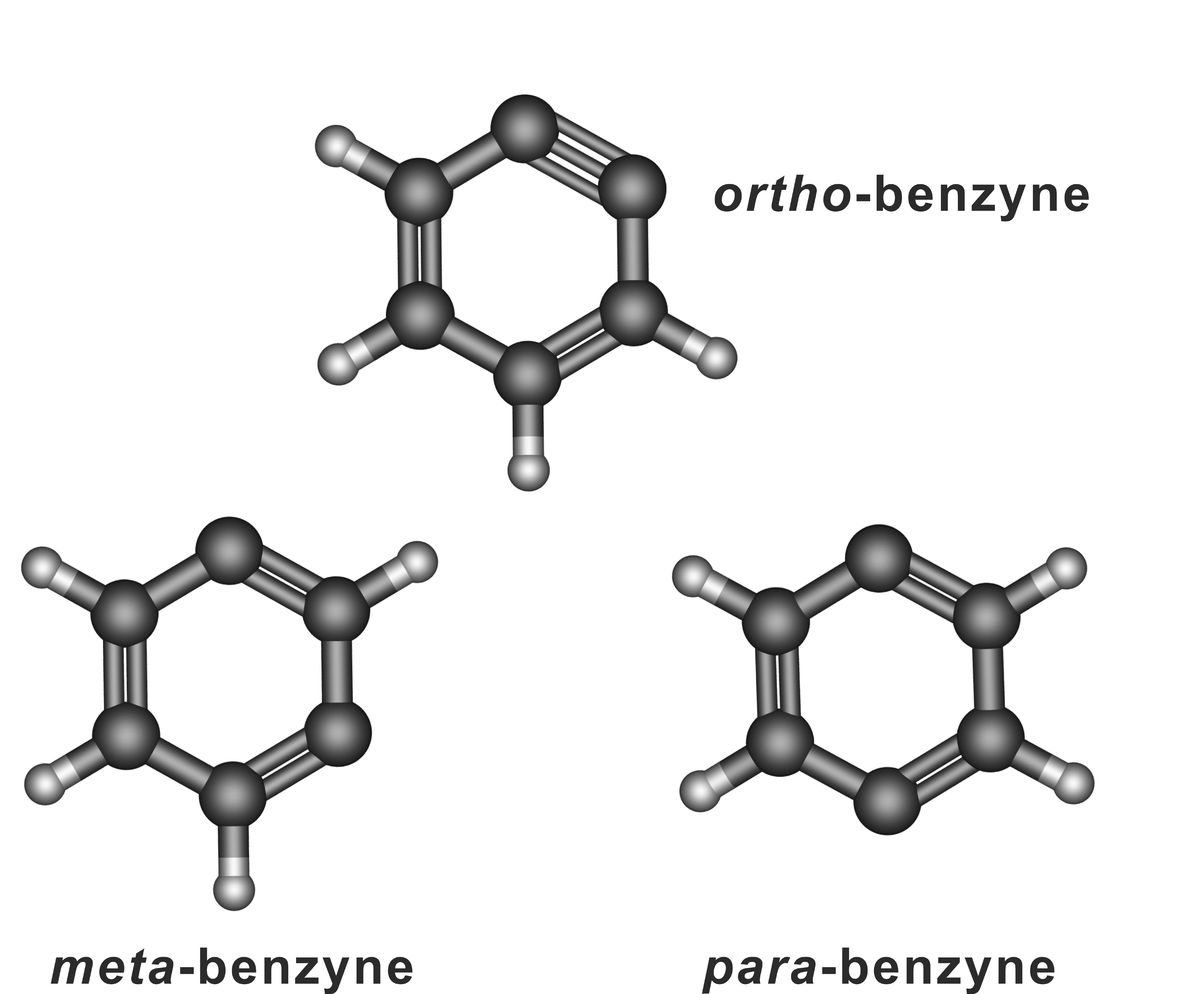}
\caption{Scheme of the ortho, para, and meta isomers of benzyne.}
\label{fig_structure_benzyne}
\end{figure}

The name benzyne refers to three isomeric species with molecular formula 
$c$-C$_6$H$_4$, which are six-membered rings with four hydrogen atoms. They can 
be considered as derivatives of benzene molecules due to the loss of two hydrogen 
atoms. The most stable isomer is 1,2-didehydrobenzene (ortho-benzyne), followed 
by 1,3-didehydrobenzene (meta-benzyne) and the 1,4-didehydrobenzene (para-benzyne), 
placed at 147 and 220 kJ mol$^{-1}$ from the ortho-benzyne, respectively \citep{Olsen1971}.
The molecular structure of ortho-benzyne (hereinafter referred to as benzyne or as 
$o$-C$_6$H$_4$)
is interesting because it contains a formal triple bond (`-yne') within a six-membered ring 
(see Fig. \ref{fig_structure_benzyne}). 
This peculiar structure confers a high reactivity, and thus it is a reactive intermediate in 
organic chemistry that plays an important role in aromatic nucleophilic substitution reactions. 
In addition, it is a potential key species in the formation of PAHs and soot in 
pyrolysis experiments 
of organic molecules \citep{Hirsch2018}.

In this letter we report the discovery of the ortho isomer of benzyne, $o$-C$_6$H$_4$, 
and we discuss
the relative abundances of pure hydrocarbons, chains and cycles, in the context of chemical models
that include new radical-neutral reactions involving the abundant hydrocarbons recently found
in TMC-1 \citep{Agundez2021a,Cernicharo2021a,
Cernicharo2021b,Cernicharo2021c}. We also discuss the evolution of these 
molecules when the volume density increases due to the gravitational collapse of the core to 
form protostars and their dusty planetary disks.

\section{Observations}
\label{observations}
New receivers, built within the Nanocosmos project\footnote{\texttt{https://nanocosmos.iff.csic.es/}}
and installed at the Yebes 40m radio telescope, were used
for the observations of TMC-1
($\alpha_{J2000}=4^{\rm h} 41^{\rm  m} 41.9^{\rm s}$ and $\delta_{J2000}=
+25^\circ 41' 27.0''$). A detailed description of the system is 
given by \citet{Tercero2021}.
The receiver consists of two cold high electron mobility transistor amplifiers that cover the
31.0-50.3 GHz band with horizontal and vertical             
polarisations. Receiver temperatures in the runs achieved in 2020 vary from 22 K at 32 GHz
to 42 K at 50 GHz. Some power adaptation in the down-conversion chains have reduced
the receiver temperatures in 2021 to 16\,K at 32 GHz and 25\,K at 50 GHz.
The back ends are $2\times8\times2.5$ GHz fast Fourier transform spectrometers
with a spectral resolution of 38.15 kHz,
providing the whole coverage of the Q band in both polarisations. 
All observations were performed in the frequency 
switching mode. The main beam efficiency varies from 0.6 at
32 GHz to 0.43 at 50 GHz. The intensity scale used in this work, antenna temperature
($T_A^*$),  was calibrated using two absorbers at different temperatures and the
atmospheric transmission model ATM \citep{Cernicharo1985, Pardo2001}.
Calibration uncertainties were adopted to be 10~\%.
All data were analysed using the GILDAS package\footnote{\texttt{http://www.iram.fr/IRAMFR/GILDAS}}.

The QUIJOTE line survey was performed in several sessions between 2019 and 2021.  
Previous QUIJOTE results on the detection of C$_3$N$^-$ and C$_5$N$^-$
\citep{Cernicharo2020a}, HC$_5$NH$^+$ \citep{Marcelino2020}, HC$_4$NC \citep{Cernicharo2020b}, and HC$_3$O$^+$
\citep{Cernicharo2020c} were based on two observing runs performed in November 2019 and February 2020. 
Two
different frequency coverages were achieved during these runs, 31.08-49.52 GHz and 31.98-50.42 GHz, 
in order to check that no
spurious spectral ghosts are produced in the down-conversion chain. In these observations the
frequency throw was 10 MHz.

Additional data were taken in October 2020, December 2020, and January-May 2021. 
In these observations the selected frequency coverage was 31.08-49.52 GHz, and the frequency
throw for the frequency switching observations was 8 MHz. 
The two different frequency throws allow unambiguously negative features
produced in the folding of the frequency switching data to be identified.
These data allowed the detection
of the acetyl cation, CH$_3$CO$^+$ \citep{Cernicharo2021d}, HC$_3$S$^+$ \citep{Cernicharo2021e},
the isomers of C$_4$H$_3$N \citep{Marcelino2021}, $l$-H$_2$C$_5$ \citep{Cabezas2021c}, vinyl and allenyl 
acetylene \citep{Cernicharo2021a,Cernicharo2021b}, ethynyl cyclopropenylidene, cyclopentadiene, and
indene \citep{Cernicharo2021c}. Several sulphur-bearing species (NCS, HCCS, H$_2$CCS, H$_2$CCCS, 
C$_4$S, HCSCN, and HCSCCH) have also been detected with
these data \citep{Cernicharo2021f,Cernicharo2021g}. In addition, some molecules typical of hot cores
and corinos (C$_2$H$_3$CHO, C$_2$H$_3$OH, HCOOCH$_3$, CH$_3$OCH$_3$, CH$_3$CH$_2$CN) 
and species typical of circumstellar envelopes (C$_5$S, HCCN, HC$_4$N)
were also detected \citep{Agundez2021b,Cernicharo2021a}.
Last but not least, a significant number of singly and doubly deuterated species, such
as HDCCN \citep{Cabezas2021a}, CH$_2$DC$_3$N \citep{Cabezas2021b}, and the D and $^{13}$C doubly 
substituted isotopologues
of HC$_3$N (Tercero et al. 2021, in preparation),
have also been observed for the first time in space thanks to QUIJOTE.

The total observing
time on the source by May 2021 was 240 hours. QUIJOTE is a living line survey, and we 
expect a total observing
time on the source of 450 hours by the end of 2021. The final goal of QUIJOTE, a 
sensitivity of 0.1 mK (lines of 0.5 mK detected at the 5$\sigma$ level), will require around 1000 hours
of observing time and will be reached by the winter of 2023-2024.

The sensitivity of our TMC-1 data is better than
that of previously published line surveys of this source at the same frequencies \citep{Kaifu2004} by a factor 20-30. 
Below 40 GHz the sensitivity is 0.2 mK, increasing slowly up to 0.5 mK at 49.5 GHz.
Thanks to this achievement, it has been possible to detect many individual lines from 
molecules \citep{Cernicharo2021a}
that have been reported only by stacking techniques \citep{Burkhardt2021}. The level of sensitivity we
have reached transforms TMC-1 into a source in which, although still far from the spectral confusion limit,
the density of lines below 1 mK in intensity is particularly high. TMC-1 can no longer be considered as a line-poor source. Hence, the identification of new species requires a systematic exploration of the
rotational transitions of single and double isotopologues of abundant species 
(see e.g. \citealt{Cabezas2021a,Cabezas2021b}, Tercero et al. 2021, in preparation)
before the detection of a new species can be claimed. In this context, line identification in this work was done using the catalogues 
MADEX \citep{Cernicharo2012}, CDMS \citep{Muller2005}, JPL \citep{Pickett1998}, 
and Splatalogue\footnote{https://splatalogue.online/advanced1.php}.
By May 2021, the MADEX code contained 6369 spectral
entries corresponding to the ground and vibrationally excited states, together
with the corresponding isotopologues, of 1691 molecules.

\begin{figure*}
\centering
\includegraphics[scale=0.85]{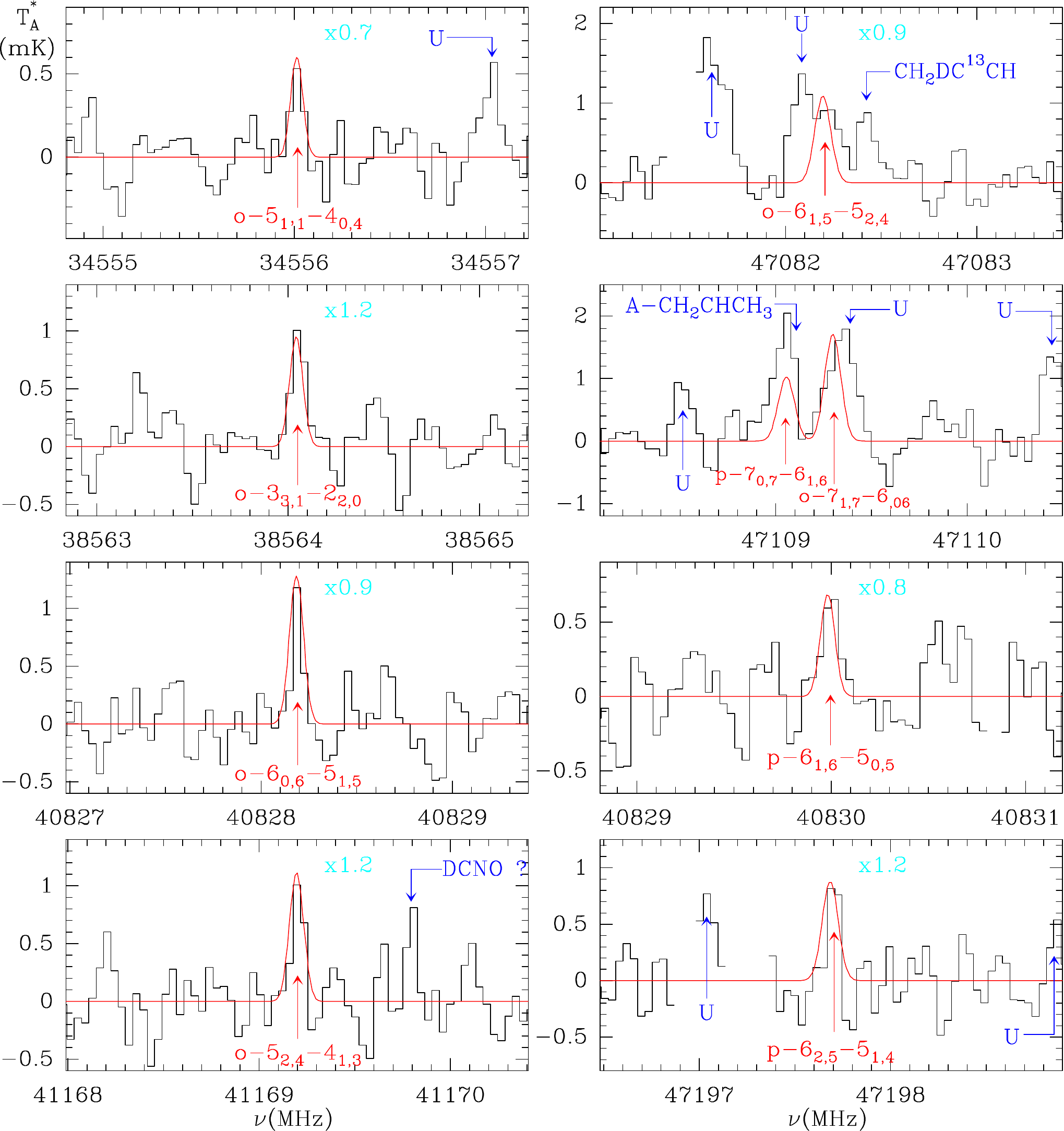}
\caption{Subset of the observed lines of $o$-C$_6$H$_4$ in the 31-50 GHz 
frequency range towards TMC-1 (see Table \ref{line_parameters} for the 
complete list of detected benzyne lines).
The abscissa corresponds to the rest frequency assuming a local standard of rest velocity of 5.83
km s$^{-1}$. 
The ordinate is the antenna temperature corrected for atmospheric and telescope losses, in mK.
The red line shows the synthetic spectrum obtained from a fit to the observed line profiles,
which provides T$_r$=7$\pm$1\,K and N($o$-C$_6$H$_4$)=(5.0$\pm$1.0)$\times$\once. 
The rotational quantum
numbers are indicated in each panel. Blanked channels correspond to negative features produced
in the folding of the frequency switching data. Cyan numbers indicate the multiplicative 
factor applied to the best model, when needed, to match the observations.}
\label{fig_c6h4}
\end{figure*}

\section{Results}
\label{results}

Our recent discovery of pure hydrocarbon cycles such as cyclopentadiene and 
indene \citep{Cernicharo2021c} prompted us to search for other chemically 
related hydrocarbon cycles, benzyne ($o$-C$_6$H$_4$) being an obvious search candidate.
The benzyne molecule is an asymmetric top with a $b$ dipole directed along the C$_{2v}$ 
symmetry axis, estimated from ab initio calculations to be 1.38\,D \citep{Kraka1993}. 
Laboratory data for this species have been taken from \citet{Brown1986}, \citet{Kukolich2003},
and \citealt{Robertson2003}.
The rotational levels of benzyne are subjected to spin statistics due to the 
presence of two pairs of equivalent hydrogen nuclei. As shown in \citet{Bunker1998}, 
the nuclear statistical weight for the rotational levels with $K_a+K_c$ even and odd is five and three,
respectively (ortho and para species; these names refer to the symmetry states and
are unrelated to the same names used to distinguish the different isomers). The 
first para level, $1_{0,1}$, is 0.42 K above the ground ortho state, $0_{00}$. 
Ortho-benzyne
was previously searched for towards CRL618 without success \citep{Weaver2007}.

An examination of the QUIJOTE data quickly indicates that all ortho lines are detected
with antenna temperatures between 0.5 and 1.8 mK. A total of seven ortho
lines and four
para lines are detected. Figure \ref{fig_c6h4} shows the nine lines detected above
3$\sigma$. Two of the lines of para benzyne are detected 
only at the 3$\sigma$ level and are not shown in this figure
(see Table \ref{line_parameters}). 
Nevertheless, their intensities are compatible with our model predictions. 
Table \ref{line_parameters}
provides the line parameters of the observed lines. One of the para lines, the
$3_{3,0}-2_{2,1}$, is fully blended with a line of 2 mK arising from 
D$^{13}$CCCN (Tercero et al. 2021, in preparation). Three additional lines of the para species are below the
3$\sigma$ detection limit at the corresponding frequencies (see Table \ref{line_parameters}).
Some interfering features produce partial blends with three lines of benzyne, as shown in Fig. \ref{fig_c6h4}.
One of these features arises from a rotational transition of the $A$ species of propylene and
overlaps in frequency with the $7_{0,7}-6_{1,6}$ transition of benzyne. 
From the derived column density of propylene by \citet{Marcelino2007}, the expected
contribution from the transition of benzyne is 0.5\,mK.
A couple of unidentified lines
also have some partial blend with the $7_{0,7}-6_{1,6}$ and $6_{1,5}-5_{2,4}$ transitions of
benzyne. However, the line
parameters for these transitions can be derived reasonably well.
All remaining lines in the Q band of benzyne are well below our present detection limit.
At the present level of sensitivity of QUIJOTE, the number of unidentified lines becomes a
concern when any stacking technique is used to detect molecules that produce weak emission. 
Our detection of benzyne
based on the observation of several ortho and para lines
is robust, reliable, and provides an accurate estimation of its column
density. 

An analysis of the observed intensities
using a line profile fitting method \citep{Cernicharo2021e} provides a rotational temperature
of 7$\pm$1\,K and a total (ortho plus para) column density for benzyne of (5.0$\pm$1.0)$\times$\once.
In the fitting procedure we considered the ortho and para levels as those of 
a single molecule with the corresponding statistical weights (5/3). 
A separate fit considering the two symmetry species separately and 
fixing the rotational temperature to 7\,K 
provides similar results and an ortho/para ratio of 1.4$\pm$0.6, which is consistent
with the expected value of 5/3.

Adopting an H$_2$ column density of 10$^{22}$ cm$^{-2}$ for TMC-1 \citep{Cernicharo1987}, the 
abundance of $o$-C$_6$H$_4$ relative
to H$_2$ is (5.0$\pm$1)$\times$\,10$^{-11}$. This is significantly below those of other cycles, such as 
$c$-C$_3$H$_2$, $c$-C$_5$H$_6$,
and $c$-C$_9$H$_8$ \citep{Cernicharo2021a}.

\begin{figure}
\centering
\includegraphics[width=\columnwidth]{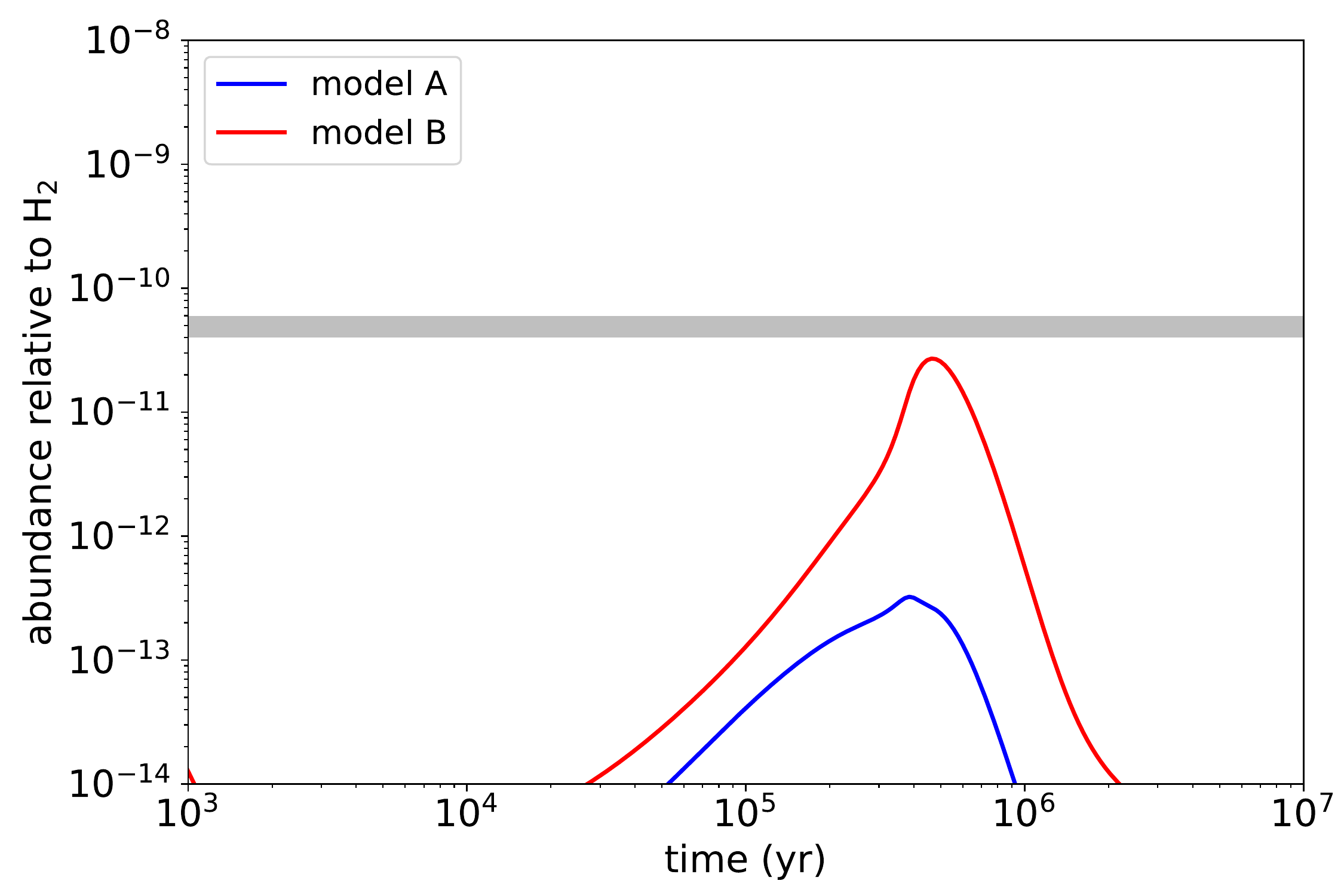}
\caption{Calculated fractional abundance of $o$-C$_6$H$_4$
as a function of time in models A and B (see text). The horizontal grey band corresponds to the observed abundance in TMC-1.}
\label{fig_abun}
\end{figure}

\section{Chemistry} \label{chemistry}

The detection of $o$-C$_6$H$_4$ in TMC-1 raises interesting questions about 
its formation. The possible reactions leading to this molecule are discussed
in Appendix \ref{network}. 
To investigate the overall contribution of reactions (1) and (7) to 
the formation $o$-C$_6$H$_4$ in TMC-1, astrochemical models were exploited 
(see Fig. \ref{fig_abun} and Appendix \ref{network}). We built a pseudo-time-dependent gas-phase 
chemical model. 
We adopted typical parameters of cold dense clouds, namely a gas kinetic 
temperature of 10 K, a volume  density of H$_2$ of 2$\times$10$^4$ cm$^{-3}$ , a cosmic-ray  
ionisation  rate of H$_2$ of 1.3$\times$10$^{-17}$  s$^{-1}$, a visual 
extinction of 30 mag, and the set of so-called `low-metal' elemental 
abundances (see e.g. \citealt{Agundez2013}). We note that the volume density of 
H$_2$ in TMC-1 
could be somewhat higher than that in our adopted value. The main effect of a slight 
increase in the density is that the abundance curves shift to shorter times, 
while peak abundances may increase or decrease slightly. The gas-phase 
chemical  network  is based on the RATE12 network from the UMIST database 
\citep{McElroy2013}, with updates relevant for the chemistry of CH$_2$CCH, 
CH$_2$CHCCH, H$_2$CCCHCCH, and $c$-C$_5$H$_6$ \citep{Agundez2021a,Cernicharo2021a,
Cernicharo2021b,Cernicharo2021c}. Benzyne was included and assumed to be 
formed by reactions (1) and (7) and destroyed by reactions with C atoms 
and C$^+$ and H$^+$  ions.  
The peak fractional abundance of $o$-C$_6$H$_4$ is 3$\times$10$^{-13}$ 
relative to H$_2$ (see model A in Fig.~\ref{fig_abun}), 
which is well below the observed value, (5$\pm$1)$\times$10$^{-11}$. This 
suggests that alternative reactions 
besides (1) and (7) may contribute to $o$-C$_6$H$_4$.

Among these reactions, (2), (4), (5), (8), (9), (11), and (12) 
are unlikely to form $o$-C$_6$H$_4$ (Appendix \ref{network}). 
Reactions (3) and (6) are 
difficult to implement since the radicals C$_5$H$_3$ and C$_4$H$_5$ are not 
included in the UMIST database. Adding 
ill-defined 
radical reactions would add more uncertainty to the output of the models. 
However, the remaining reactions, (13) and (14) and potentially (10), are 
plausible routes to $o$-C$_6$H$_4$. Moreover, these reactions involve reactants 
that are abundant in TMC-1 \citep{Fosse2001,Cabezas2021a,Agundez2021a}. Including these 
three reactions results in a peak abundance 
for $o$-C$_6$H$_4$ of 
3$\times$10$^{-11}$ relative to H$_2$ (model B in Fig.~\ref{fig_abun}), which is 
in good agreement  
with the observed value, (5$\pm$1)$\times$10$^{-11}$. Therefore, 
reactions (13), (14), and (10) deserve to be studied experimentally and theoretically in the future to understand how this highly reactive 
cyclic and aromatic molecule, $o$-C$_6$H$_4$, can be formed in TMC-1. It should be noted that 
benzyne could also be formed through ionic routes. For example, the 
dissociative recombination of the $c$-C$_6$H$_5^+$ ion with electrons could 
produce 
$o$-C$_6$H$_4$. 
The precursor ion, $c$-C$_6$H$_5^+$, could in turn be formed through different ion-neutral reactions, such as 
C$_4$H$_3^+$ + C$_2$H$_2$ (\citealt{Anicich1993}; 
see also \citealt{Woods2002}) or C$_3$H$_3^+$ + CH$_2$CCH.

It would be interesting to investigate whether $o$-C$_6$H$_4$ could be 
formed upon the dissociative recombination of C$_6$H$_5^+$ or C$_6$ ions 
with more than five H atoms.

\section{Discussion and conclusions}
The high abundance of hydrocarbon chains and cycles in TMC-1 
suggests that one of the main sites of production of the seeds 
for PAH formation could be cold dark clouds and that this process could 
be a transitory one. Figure \ref{fig_abun} shows that the peak of formation of 
$o$-C$_6$H$_4$ and other hydrocarbons is produced in a few 10$^5$ years. When the 
density increases due to the collapse of the cloud, most of these 
large hydrocarbons will stick onto dust grains. The chemistry of 
ices of cyclopentadiene, benzene, indene, and other species is poorly 
known as most experiments have been done with large PAHs. The dust 
grains enriched in these molecules, together with their pristine 
carbon composition produced during their early formation, will 
evolve with time and will become the constituents of the solid blocks 
that produce asteroids and planetesimals during the first stages
of star and planet formation. Eventually, some of these grains 
will be exposed to UV radiation from external and/or internal sources 
producing photodissociation 
regions. A significant fraction of the PAH emission observed 
in these regions could be due to PAHs formed from hydrocarbon-rich ices 
\citep{Abplanalp2018,Kaiser1997}, and ices of
cyclopentadiene, benzene, and indene deposited in a transitory 
phase of between 10$^5$ and 10$^6$ yr during the evolution of the cloud 
(we note that chemical times depend on the volume density and adopted initial 
physical and chemical conditions).
None of these species have been found in the deep Q-band line 
survey of IRC\,+10216 performed with the Yebes 40m radio telescope \citep{Pardo2021}. 
While in the external layers of this carbon-rich circumstellar envelope the abundance of radicals is very high
\citep{Agundez2017}, the chemical time corresponding to these regions is 
relatively short, $\sim$10$^4$ yr, implying that the first cycles may not have had enough time to be formed. A different context is that 
of carbon-rich protoplanetary nebulae such as CRL\,618, where photodissociation and ionisation is
enhanced with respect to the AGB phase and where benzene has been found from infrared observations
with the Infrared Space Observatory \citep{Cernicharo2001}.
Further, distinct temperature ranges from 10 K to a few thousand K facilitate 
diverse low- and high-temperature molecular mass growth processes 
\citep{Cernicharo2004,Kaiser2021}.

To summarise, we report in this work the detection of the ortho isomer 
of benzyne, $o$-C$_6$H$_4$, in TMC-1. The observed abundance, 
(5$\pm$1)$\times$10$^{-11}$ 
relative to H$_2$ , can be nicely reproduced with fractional abundances 
of 3$\times$10$^{-11}$ considering that $o$-C$_6$H$_4$ is 
formed by the neutral-neutral 
reactions C$_2$H + CH$_2$CHCCH and C + C$_5$H$_5$ and possibly through 
the reactions  C$_4$H + C$_2$H$_4$, C$_3$H + CH$_2$CCH$_2$, 
and C$_3$H$_2$ + C$_3$H$_3$.  With a sensitive  
line survey such as QUIJOTE,  the discovery of new molecules can be 
performed in the classical line by line detection method. Hence, QUIJOTE  
provides unambiguous detection of new molecular species and reliable 
column density determinations.

\begin{acknowledgements}
We thank ERC for funding
through grant ERC-2013-Syg-610256-NANOCOSMOS. M.A. thanks MICIU for grant 
RyC-2014-16277. We also thank Ministerio de Ciencia e Innovaci\'on of Spain (MICIU) for funding support through projects
AYA2016-75066-C2-1-P, PID2019-106110GB-I00, PID2019-107115GB-C21 / AEI / 10.13039/501100011033, 
and PID2019-106235GB-I00. 

\end{acknowledgements}

\begin{appendix}
\section{Benzyne line parameters}
Line parameters were derived from a Gaussian fit to the observed lines using
the GILDAS package. A velocity
coverage of $\pm$15\,\kms\, was selected for each line. Observed frequencies were derived assuming a local standard of rest velocity of 5.83\kms \citep{Cernicharo2020b}. 
Observed velocities for the detected lines 
were determined assuming that rest frequencies are those predicted from the
rotational and distortion constants provided by the fit to the laboratory data 
\citep{Brown1986,Kukolich2003,Robertson2003}. The predicted frequencies in MADEX \citep{Cernicharo2012}
agree within 1-5 kHz with those of the CDMS \citep{Muller2005}. Differences between predicted and
observed frequencies never exceed 1.5$\sigma\Delta\nu$, where $\Delta\nu$ is 
the estimated frequency uncertainty of the observed lines. The averaged V$_{LSR}$ from the 11
observed lines of benzyne is 5.80$\pm$0.05\kms, a value similar to that derived from all lines of
HC$_5$N and its isotopologues in the Q band \citep{Cernicharo2020b}.

\begin{table*}
\centering
\caption{Observed line parameters of benzyne.}
\label{line_parameters}
\begin{tabular}{lccccccr}
\hline
Transition       & $\nu_{pred}$\,$^a$ & $\nu_{obs}\,^b$ &  $\int$ $T_A^*$ dv $^c$ & v$_{LSR}$\,$^d$   &$\Delta$v\,$^e$ & $T_A^*$\,$^f$ & N  \\
                 & (MHz)         & (MHz)        & (mK\,km\,s$^{-1}$)   & (km\,s$^{-1}$)& (km\,s$^{-1}$)  & (mK)   &  \\
\hline
ortho state      &             &              &         &         &         &           &  \\
$4_{1,3}-3_{2,2}$& 33482.756(1)& 33482.786(30)& 0.34(07)& 5.56(07)& 0.66(18)& 0.48(17)  &  \\
$5_{1,5}-4_{0,4}$& 34556.013(2)& 35556.013(30)& 0.36(10)& 5.83(09)& 0.63(18)& 0.55(18)  &  \\
$3_{3,1}-2_{2,0}$& 38564.042(1)& 38564.054(30)& 0.74(15)& 5.74(07)& 0.68(17)& 1.02(20)  &  \\
$6_{0,6}-5_{1,5}$& 40828.169(4)& 40828.169(20)& 0.54(11)& 5.83(05)& 0.45(10)& 1.19(22)  &  \\
$5_{2,4}-4_{1,3}$& 41169.208(2)& 41169.205(20)& 0.56(13)& 5.85(07)& 0.48(12)& 1.09(24)  &  \\
$6_{1,5}-5_{2,4}$& 47082.207(4)& 47082.207(30)& 0.87(30)& 5.83(16)& 0.87(30)& 0.94(27)  &  \\
$7_{1,7}-6_{0,6}$& 47109.298(7)& 47109.345(30)& 1.91(25)& 5.53(10)& 1.02(15)& 1.77(35)  & A\\
\hline                                                                           
para state       &             &              &         &         &         &           &  \\
$5_{0,5}-4_{1,4}$& 34543.093(2)&              &         &         &         & $\le$0.70 &  \\
$4_{2,3}-3_{1,2}$& 35656.830(1)&              &         &         &         & $\le$0.75 &  \\
$5_{1,4}-4_{2,3}$& 40595.926(2)& 40595.919(30)& 0.16(08)& 5.90(12)& 0.40(15)& 0.47(17)  &  \\
$6_{1,6}-5_{0,5}$& 40829.994(4)& 40830.000(30)& 0.55(19)& 5.79(13)& 0.78(39)& 0.67(27)  &  \\
$3_{3,0}-2_{2,1}$& 41071.944(1)&              &         &         &         &           & B\\
$4_{3,2}-3_{2,1}$& 45028.539(1)&              &         &         &         & $\le$0.75 &  \\
$7_{0,7}-6_{1,6}$& 47109.056(7)& 47109.044(20)& 1.63(21)& 5.90(08)& 0.80(13)& 1.92(27)  & C\\
$6_{2,5}-5_{1,4}$& 47197.685(4)& 47197.708(30)& 0.45(11)& 5.72(20)& 0.45(15)& 0.95(26)  &  \\
\hline
\end{tabular}
\tablefoot{\\
Values between parentheses correspond to the uncertainties of the parameters 
in units of the last significant digits.\\
Upper limits correspond to 3$\sigma$ values.\\
\tablefoottext{a}{Predicted frequency from the rotational and distortion constants
derived from a fit to the lines observed by \citet{Brown1986}, \citet{Kukolich2003},
and \citet{Robertson2003}.}\\
\tablefoottext{b}{Observed frequency assuming a v$_{LSR}$ of 5.83 \kms.}\\
\tablefoottext{c}{Integrated line intensity in mK\,km\,s$^{-1}$.}\\
\tablefoottext{d}{v$_{LSR}$ assuming that the predicted frequencies are the rest frequencies of the lines (in km\,s$^{-1}$).}\\
\tablefoottext{e}{Line width at half intensity derived by fitting a Gaussian function to
the observed line profile (in km\,s$^{-1}$).}\\
\tablefoottext{f}{Antenna temperature in millikelvin.}\\
\tablefoottext{A}{Possibly a blend with a U line at +35 kHz of the predicted frequency.}\\
\tablefoottext{B}{Blended with a line of D$^{13}$CCCN. Fit unreliable.}\\
\tablefoottext{C}{Blended with a line of CH$_2$CHCH$_3$ with an expected intensity of 0.5 mK.}\\
}
\end{table*}

\section{The chemical network for the formation of benzyne}
\label{network}
The low number densities of the reactants along with the typical 
10 K translational temperature of a cold starless core such as TMC-1
only support barrierless and exoergic bimolecular reactions. Focusing on 
neutral-neutral reactions, the scheme is 
A + B $\rightarrow$[AB]$^*$ $\rightarrow$ C + D, where A and B represent 
the reactants, [AB]* is the reaction intermediate(s), and C and D are the products 
\citep{Kaiser2002}. In TMC-1, 
the newly detected $o$-C$_6$H$_4$ defines the product C; atomic hydrogen 
represents one of the most prominent counter fragments, D, that carry away a 
significant amount of the total angular momentum. This concludes that 
[AB]$^*$ has to fulfil the stoichiometry of [C$_6$H$_5$]$^*$. Since the intermediate 
holds six carbon atoms, this can be achieved in bimolecular reactions 
in C$_1$-C$_5$, C$_2$-C$_4$, and C$_3$-C$_3$ systems with reactants at 
various degrees of hydrogenation. In principle, the following reactions, 
(1) to (14), could yield rovibrationally excited [C$_6$H$_5$]$^*$ reaction 
intermediates, which undergo unimolecular decomposition via atomic 
hydrogen loss to form the newly detected $o$-C$_6$H$_4$ molecule.

(1) C + C$_5$H$_5$           $\rightarrow$ $o$-C$_6$H$_4$ + H

(2) CH + C$_5$H$_4$          $\rightarrow$ $o$-C$_6$H$_4$ + H

(3) CH$_2$ + C$_5$H$_3$      $\rightarrow$ $o$-C$_6$H$_4$ + H

(4) CH$_3$ + C$_5$H$_2$      $\rightarrow$ $o$-C$_6$H$_4$ + H

(5) CH$_4$ + C$_5$H          $\rightarrow$ $o$-C$_6$H$_4$ + H

(6) C$_2$ + C$_4$H$_5$       $\rightarrow$ $o$-C$_6$H$_4$ + H

(7) C$_2$H + C$_4$H$_4$      $\rightarrow$ $o$-C$_6$H$_4$ + H

(8) C$_2$H$_2$ + C$_4$H$_3$  $\rightarrow$ $o$-C$_6$H$_4$ + H

(9) C$_2$H$_3$ + C$_4$H$_2$  $\rightarrow$ $o$-C$_6$H$_4$ + H

(10) C$_2$H$_4$ + C$_4$H     $\rightarrow$ $o$-C$_6$H$_4$ + H

(11) C$_2$H$_5$ + C$_4$      $\rightarrow$ $o$-C$_6$H$_4$ + H

(12) C$_3$ + C$_3$H$_5$      $\rightarrow$ $o$-C$_6$H$_4$ + H                   

(13) C$_3$H + C$_3$H$_4$     $\rightarrow$ $o$-C$_6$H$_4$ + H

(14) C$_3$H$_2$ + C$_3$H$_3$ $\rightarrow$ $o$-C$_6$H$_4$ + H.

Among these reactions, only the reaction between the ethynyl 
radical (C$_2$H) and vinylacetylene (CH$_2$CHCCH), reaction (7), has been 
studied via crossed molecular beam experiments and electronic 
structure calculations \citep{Zhang2011}. This reaction is 
barrierless and exoergic, and thus it is expected to be fast at 
low temperatures. This finding is in line with the known high 
reactivity of C$_2$H radicals with unsaturated hydrocarbons at 
low temperatures \citep{Chastaing1998}, revealing rates of 
up to 4$\times$10$^{-10}$ cm$^3$s$^{-1}$.  Moreover, \citet{Zhang2011} 
predicted that $o$-C$_6$H$_4$ is formed with a branching ratio of 
up to 20\% at 10 K, thus providing an effective rate of 
8$\times$10$^{-11}$ cm$^3$s$^{-1}$. Reaction (7) is thus a plausible route 
to benzyne in TMC-1. Further, in analogy to the 
C($^3$P)–benzene system \citep{Hahndorf2002}, 
the reaction of atomic carbon (C($^3$P)) with cyclopropadienyl 
($c$-C$_5$H$_5$) – although studied neither experimentally nor 
computationally – is expected to form $o$-C$_6$H$_4$ plus atomic 
hydrogen (reaction (1)). \citet{Hahndorf2000} predicted 
computationally that the propargyl radical (C$_3$H$_3$) can react 
barrierlessly and exoergically with the vinyl radical (C$_2$H$_3$) 
on the singlet surface to form $c$-C$_5$H$_5$ plus atomic hydrogen, 
thus providing the $c$-C$_5$H$_5$ required for reaction (1). 

\subsection{Discounted reactions to benzyne}
\label{discounted}
Reaction (2) has never been studied in the laboratory with any isomer
such as methyldiacetylene (HCCCCCH$_3$), penta-1,4-diyne (CH$_2$(C$_2$H)$_2$), 
or 1,2,3,4-pentatetraene (H$_2$CCCCCH$_2$). Rate constant measurements of 
methylidyne (CH) with unsaturated hydrocarbons by \citet{Canosa1997}  
reveal barrierless pathways with rates of a few 10$^{-10}$ cm$^3$s$^{-1}$.  
However, crossed molecular beam reactions indicate that the CH radical 
adds to the double or triple bond of, for example, acetylene, ethylene, diacetylene, 
allene, and methylacetylene, followed by isomerisation through the ring opening 
and/or hydrogen shifts, followed by hydrogen atom losses. Six-membered 
ring formation only occurs through the ring expansion of an existing 
five-membered ring (i.e. a hypothetical cyclic $c$-C$_5$H$_4$ isomer). 
However, no confirmed formation routes to cyclic $c$-C$_5$H$_4$ are known 
so far; therefore, we do not include reaction (2) as a contributor to benzyne. 

Reaction (4), which has never been studied either computationally nor 
experimentally, involves the addition of a methyl radical (CH$_3$) to 
a triplet carbene (HCCCCCH). The addition of a methyl radical to triple 
bonds of the ethynyl moiety involves barriers of up to 50 kJ mol$^{-1}$,  
which cannot be overcome in TMC-1. Methyl radicals could add 
barrierlessly to the central carbon atom, which essentially carries 
the unpaired electrons. However, this requires small, nearly zero-impact 
parameters and an extensive reorganisation of the carbon skeleton; therefore, 
although exoergic by 207  kJ mol$^{-1}$ overall,  the formation of benzyne 
is unlikely. 

Reaction (5) represents a direct reaction that leads, via hydrogen 
abstraction, to a methyl radical plus C$_5$H$_2$ molecules; although 
never studied experimentally, this hydrogen abstraction is expected 
to have a barrier. Further, this reaction is endoergic by 27 kJ mol$^{-1}$ 
and hence cannot take place at 10 K. Reaction (5) cannot form 
$o$-C$_6$H$_4$ + H at 10 K since hydrogen abstraction represents the only 
feasible pathway.

Reaction (8) has an entrance barrier of addition of the 
C$_4$H$_3$  with its radical centre to the carbon-carbon triple bond 
of acetylene  25–37 kJ mol$^{-1}$. Hence, its reaction is closed at 10 
K as well.

Reaction (9) involves the reaction of a vinyl radical with diacetylene; 
reactions of vinyl radicals with closed-shell olefines and alkynes have 
entrance barriers of 10 to 35 kJ mol$^{-1}$, which cannot be overcome in TMC-1. 
This has been confirmed computationally (Zhang et al. 2011).  Finally, 
reaction (11) and, in particular, reaction (12) likely have entrance barriers 
\citep{Mebel2015}.

\end{appendix}

\begin{thebibliography}{} \tiny
\bibitem[Abplanalp et al.(2018)]{Abplanalp2018}Abplanalp, M.J., Jones, B.M., Kaiser, R.I. 2018, PCCP, 20, 5435
\bibitem[Ag\'undez \& Wakelam(2013)]{Agundez2013} Ag\'undez, M. \& Wakelam, V. 2013, Chem. Rev. 113, 8710
\bibitem[Ag\'undez et al.(2017)]{Agundez2017} Ag\'undez, M. Cernicharo, J., Quintana-Lacaci, G., et al. 2017, \aap, 601, A4
\bibitem[Ag\'undez et al.(2021a)]{Agundez2021a} Ag\'undez, M., Cabezas, C., Tercero, B., et al. 2021a, \aap, 647, L10 
\bibitem[Ag\'undez et al.(2021b)]{Agundez2021b} Ag\'undez, M., Marcelino, N., Tercero, B., et al. 2021b, \aap, 649, L4 
\bibitem[Anicich(1993)]{Anicich1993} Anicich, V. G. 1993, J. Phys. Chem. Ref. Data, 22, 1469
\bibitem[Berteloite et al.(2010)]{Berteloite2010} Berteloite, C., Le Picard, S. D., Balucani, N., et al. 2010, PCCP, 12, 3677
\bibitem[Bunker \& Jensen (1998)]{Bunker1998}Bunker, P.R. \& Jensen, P. 1998, Molecular Symmetry and Spectroscopy, 2nd edn. (Ottawa: NRC Research Press)
\bibitem[Burkhardt et al.(2021)]{Burkhardt2021}Burkhardt, A.M., Lee, L.K., Bryan Changala, P., et al. 2021, \apj, 913, L18
\bibitem[Brown et al. (1986)]{Brown1986}Brown, R.D., Godfrey, P.D. \& Rodler, M. 1986, J. Am. Chem. Soc. 108, 1296
\bibitem[Cabezas et al.(2021a)]{Cabezas2021a} Cabezas, C., Endo, Y., Roueff, E., et al. 2021a, \aap, 646, L1 
\bibitem[Cabezas et al.(2021b)]{Cabezas2021b} Cabezas, C., Roueff, E., Tercero, B., et al. 2021b, \aap, 650, L15 
\bibitem[Cabezas et al.(2021c)]{Cabezas2021c} Cabezas, C., Tercero, B., Ag\'undez, M., et al. 2021c, \aap, 650, L9 
\bibitem[Canosa et al.(1997)]{Canosa1997} Canosa, A., Sims, I. R., Travers, D., et al. 1997, \aap, 323, 644
\bibitem[Cernicharo(1985)]{Cernicharo1985} Cernicharo, J. 1985, Internal IRAM report (Granada: IRAM)
\bibitem[Cernicharo \& Gu\'elin(1987)]{Cernicharo1987} Cernicharo, J. \& Gu\'elin, M. 1987, \aap, 176, 299
\bibitem[Cernicharo et al.(2001)]{Cernicharo2001} Cernicharo, J., Heras, A.M., Tielens, A.G.G.M., et al. 2001, \apj, 546, L123
\bibitem[Cernicharo(2004)]{Cernicharo2004} Cernicharo, J. 2004, \apj, 608, L41
\bibitem[Cernicharo(2012)]{Cernicharo2012} Cernicharo, J., 2012, in ECLA 2011: Proc. of the European Conference on Laboratory Astrophysics, EAS Publications Series, 2012, Ed.: C. Stehl, C. Joblin, \& L. d'Hendecourt (Cambridge: Cambridge Univ. Press),
251; \texttt{https://nanocosmos.iff.csic.es/?page$\_$id=1619}
\bibitem[Cernicharo et al.(2020a)]{Cernicharo2020a} Cernicharo, J., Marcelino, N., Pardo, J.R., et al. 2020a, \aap, 641, L9 
\bibitem[Cernicharo et al.(2020b)]{Cernicharo2020b} Cernicharo, J., Marcelino, N., Ag\'undez, et al. 2020b, \aap, 642, L8 
\bibitem[Cernicharo et al.(2020c)]{Cernicharo2020c} Cernicharo, J., Marcelino, N, Ag\'undez, M., et al. 2020c, \aap, 642, L17 
\bibitem[Cernicharo et al.(2021a)]{Cernicharo2021a} Cernicharo, J., Ag\'undez, M., Cabezas, C., et al. 2021a, \aap, 647, L2 
\bibitem[Cernicharo et al.(2021b)]{Cernicharo2021b}  Cernicharo, J., Cabezas, C., Ag\'undez, M., et al. 2021b, \aap, 647, L3 
\bibitem[Cernicharo et al.(2021c)]{Cernicharo2021c} Cernicharo, J., Ag\'undez, M., Cabezas, C., et al. 2021c, \aap, 649, L15 
\bibitem[Cernicharo et al.(2021d)]{Cernicharo2021d} Cernicharo, J., Cabezas, C., Bailleux, S., et al. 2021d, \aap, 646, L7 
\bibitem[Cernicharo et al.(2021e)]{Cernicharo2021e} Cernicharo, J., Cabezas, C., Endo, Y., et al. 2021e, \aap, 646, L3 
\bibitem[Cernicharo et al.(2021f)]{Cernicharo2021f} Cernicharo, J., Cabezas, C., Ag\'undez, M., et al. 2021f, \aap, 648, L3 
\bibitem[Cernicharo et al.(2021g)]{Cernicharo2021g} Cernicharo, J., Cabezas, C., Ag\'undez, M., et al. 2021g, \aap, 650, L14 
\bibitem[Chastaing et al.(1998)]{Chastaing1998} Chastaing, D., James, P. L., Sims, I. R., \& Smith, I. W. M. 1998, Faraday Discussions, 109, 165
\bibitem[Cherchneff et al.(1992)]{Cherchneff1992} Cherchneff, I., Barker, J. R., \& Tielens, A. G. G. M. 1992, \apj, 401, 269
\bibitem[Kraka \& Cremer (1993)]{Kraka1993}Kraka, E. \& Cremer, D. 1993, Chem. Phys. Lett., 216, 333
\bibitem[Doddipatla et al.(2021)]{Doddipatla2021}Doddipatla, S., Galimova, G.R., Wei, H., et al. 2021, Science Advances, 7, eabd4044
\bibitem[Foss\'e et al.(2001)]{Fosse2001} Foss\'e, D., Cernicharo, J., Gerin, M., Cox, P. 2001, \apj, 552, 168
\bibitem[Hahndorf et al. (2000)]{Hahndorf2000} Hahndorf, I., Lee, Y.T., Mebel, A.M. 2000, \jcp, 113, 9622
\bibitem[Hahndorf et al. (2002)]{Hahndorf2002} Hahndorf, I., Lee, Y.T., Kaiser, R.I., et al. 2002, \jcp, 116, 3248
\bibitem[Hirsch et al. (2018)]{Hirsch2018}Hirsch, F., Reusch, E., Constantinidis, P., et al. 2018, J. Phys. Chem. A, 122, 9563
\bibitem[Joblin \& Cernicharo (2018)]{Joblin2018} Joblin, C. \& Cernicharo, J. 2018, Science, 359, 156
\bibitem[Jones et al. (2011)]{Jones2011}Jones, B.M., Zhang, F., Kaiser, R.I., et al. 2011, PNAS, 108, 452
\bibitem[Kaifu et al.(2004)]{Kaifu2004} Kaifu, N., Ohishi, M., Kawaguchi, K., et al. 2004, PASJ, 56, 69
\bibitem[Kaiser \& Roessler (1997)]{Kaiser1997} Kaiser, R.I. \& Roessler, K. 1997, \apj, 475, 144
\bibitem[Kaiser(2002)]{Kaiser2002} Kaiser, R.I. 2002, Chem. Rev., 102, 1309
\bibitem[Kaiser et al. (2012)]{Kaiser2012} Kaiser, R.I., Gu, X., Zhang, F., \& Maksyutenko, P., 2012, PCCP, 14, 575
\bibitem[Kaiser \& Hansen (2021)]{Kaiser2021} Kaiser, R.I. \& Hansen, N. 2021, J. Phys. Chem. A, 125, 3826
\bibitem[Kukolich et al. (2003)]{Kukolich2003}Kukolich, S.G., Tanjaroon, C., McCarthy, M.C. \& Thaddeus, P. 2003, \jcp, 119, 4353
\bibitem[Lide \& Mann (1957)]{Lide1957}Lide Jr, D.R. \& Mann, D.E., 1957, J. Chem. Phys., 27, 868
\bibitem[McCabe et al.(2020)]{McCabe2020}McCabe, M.N., Hemberger, P., Reusch, E., et al. 2020, J. Phys. Chem. Lett., 11, 2859
\bibitem[McElroy et al.(2013)]{McElroy2013} McElroy, D., Walsh, C., Markwick, A. J., et al. 2013, \aap, 550, A36
\bibitem[McGuire et al.(2018)]{McGuire2018} McGuire, B.A., Burkhardt, A.M., Kalenskii, S., et al. 2018, Science, 359, 202
\bibitem[McGuire et al.(2021)]{McGuire2021}McGuire, B.A., Loomis, R.A., Burkhardt, A.M., et al. 2021, Science, 371, 1265
\bibitem[Marcelino et al.(2007)]{Marcelino2007} Marcelino, N., Cernicharo, J., \& Ag\'undez, M. 2007, \apj, 665, L127
\bibitem[Marcelino et al.(2020)]{Marcelino2020} Marcelino, N., Ag\'undez, M., Tercero, B., et al. 2020, \aap, 643, L6 
\bibitem[Marcelino et al.(2021)]{Marcelino2021} Marcelino, N., Tercero, B., Ag\'undez, M., \& Cernicharo, J. 2021, \aap, 646, L9 
\bibitem[Mart\'inez et al.(2020)]{Martinez2020} Mart\'inez, L., Santoro, G., Merino, P., et al. 2020, Nature Astron., 4, 97
\bibitem[Mebel \& Kaiser (2015)]{Mebel2015}Mebel, A.M. \& Kaiser, R.I. 2015, Int. Rev. Phys. Chem., 17, 21564
\bibitem[M\"uller et al.(2005)]{Muller2005} M\"uller, H.S.P., Schl\"oder, F., Stutzki, J., Winnewisser, G. 2005, \jmst, 742, 215
\bibitem[Olsen (1971)]{Olsen1971}Olsen, J.F. 1971, \jmst, 8, 307
\bibitem[Parker et al.(2012)]{Parker2012}Parker, D.S.N., Zhang, F., Kim, Y.S., et al. 2012, PNAS, 109, 53
\bibitem[Pardo et al.(2001)]{Pardo2001} Pardo, J.~R., Cernicharo, J., Serabyn, E. 2001, IEEE Trans. Antennas and Propagation, 49, 12
\bibitem[Pardo et al.(2021)]{Pardo2021} Pardo, J.~R., et al. 2021, \aap, submitted
\bibitem[Pickett et al.(1998)]{Pickett1998} Pickett, H.M., Poynter, R.~L., Cohen, E.~A., et al. 1998, J. Quant. Spectrosc. Radiat. Transfer, 60, 883
\bibitem[Pilleri et al.(2015)]{Pilleri2015} Pilleri, P., Joblin, C., Boulanger, F. \& Onaka, T. 2015, \aap, 577, A16
\bibitem[Robertson et al. (2003)]{Robertson2003}Robertson, E.G., Godfrey, P.D. \& McNaughton, D. 2003, \jms, 217, 123
\bibitem[Tercero et al.(2021)]{Tercero2021}Tercero, F., L\'opez-P\'erez, J. A., Gallego, et al. 2021, \aap, 645, A37
\bibitem[Widicus Weaver et al. (2007)]{Weaver2007}Widicus Wwaver, S., Remijan, A.J., McMahon, R.J., and McCall, B.J. 2007, \apj, L153
\bibitem[Woods et al. (2002)]{Woods2002}Woods, P.M., Millar, T. J., \& Zijlstra, A.A. 2002, \apj, 574, L167
\bibitem[Zhang et al. (2011)]{Zhang2011} Zhang, F., Parker, D., Kim, Y. S., et al. 2011, \apj, 728, 141

\end{thebibliography}
\end{document}